\documentclass[aps]{revtex4}
\usepackage{graphicx}
\usepackage{bm}
\usepackage{graphicx}
\usepackage{amsmath}
\usepackage{mathrsfs}
\usepackage{ulem}
\usepackage{color,xcolor}
\usepackage{amsthm,amsmath,amssymb}
\usepackage[colorlinks, linkcolor=red,anchorcolor=green,citecolor=blue]{hyperref}
\usepackage{cancel}

\begin{document}

\title{Equal-time kinetic equations in a rotational field}
\author{Shile Chen}
\author{Ziyue Wang}
\email{zy-wa14@mails.tsinghua.edu.cn}
\author{Pengfei Zhuang}
\address{Physics Department, Tsinghua University, Beijing 100084, China}
\date{\today}

\begin{abstract}
We investigate quantum kinetic theory for a massive fermion system under a rotational field. From the Dirac equation in curved space we derive the complete set of kinetic equations for the spin components of the covariant and equal-time Wigner functions. While the particles are no longer on a mass shell in general case due to the rotation-spin coupling, there are always only two independent components, which can be taken as the number and spin densities. With the help from the off-shell constraint we obtain the closed transport equations for the two independent components in classical limit and at quantum level. The classical rotation-orbital coupling controls the dynamical evolution of the number density, but the quantum rotation-spin coupling explicitly changes the spin density.
\end{abstract}

\maketitle

\section{Indroduction}
From the lattice simulations~\cite{karsch} of quantum chromodynamics (QCD), it is widely accepted that there exists a phase transition from a hadron gas to a quark-gluon plasma at high temperature. The experimental efforts of high energy nuclear collisions at Relativistic Heavy Ion Collider (RHIC) and Large Hadron Collider (LHC) have provided many sensitive signatures~\cite{qm2019} of the new state of matter created in the early stage of the collisions. Considering the very short lifetime of the collision zone, the highly excited quark-gluon system spends a considerable fraction of its life in a non-equilibrium state, and the dynamical tool to treat dissipative processes in nuclear collisions and the approach to hydrodynamic evolution is in principle quantum transport theory. In classical transport theory, all the physical currents are connected with the number distribution function $f$. The quantum mechanical analogue of $f$ is the Wigner function $W$ which is a $4\times 4$ matrix in spin space~\cite{degroot}. A relativistic and gauge covariant kinetic theory for quarks and gluons has been derived, both in a classical framework~\cite{heinz} and as a quantum kinetic theory~\cite{elze} based on the Wigner functions defined in covariant~\cite{degroot} and equal-time~\cite{BialynickiBirula:1991tx} phase spaces. Many applications to the quark-gluon plasma, such as linear color response, color correlations and collective plasma oscillations~\cite{heinz2,elze2}, have been discussed in this framework using a semi-classical expansion of the quantum transport theory. 

The study on QCD phase transitions at finite temperature and density is recently extended to including electromagnetic fields, since the strongest fields in nature is believed to be generated in nuclear collisions at RHIC and LHC energies~\cite{tuchin,deng}. Under such strong electromagnetic fields some anomalous phenomena for massless quarks, such as chiral magnetic effect~\cite{Kharzeev:2007jp,Fukushima:2008xe}, are experimentally discovered in non-central nuclear collisions~\cite{star,alice}. Since the created fields drop down very fast and appear only in the very beginning of the collisions, most of the theoretical investigations is in the frame of quantum kinetic theory~\cite{Hidaka:2016yjf,Huang:2018wdl,Gao:2018wmr,Weickgenannt:2019dks,Wang:2019moi,Wang:2020pej,kharzeev}. Apart from the electromagnetic fields, the strongest rotational field in nature can also be produced in nuclear collisions. The maximum magnitude is expected to be about $0.1m_\pi$ in noncentral Au+Au collisions at RHIC energy~\cite{STAR:2017ckg,Deng:2016gyh}. A direct consequence of such strong rotation is the polarization of final state hadrons~\cite{Adam:2019srw} through spin-orbital coupling at quark level. Different from the electromagnetic fields which rapidly decay in time, the angular momentum conservation during the evolution of the collision will lead to a more visible rotational effect on the final state. The other advantage of rotational effect is that it becomes more strong in intermediate nuclear collisions where there might be new physics related to high baryon density. There have been a lot of theoretical investigations on the rotational effect and spin polarization in high energy nuclear collisions~\cite{Liang:2004ph,Becattini:2007sr,Gao:2012ix,Csernai:2013bqa,Jiang:2015cva,Becattini:2016gvu,Florkowski:2017ruc,Ivanov:2019wzg,Gao:2018jsi}. In this paper, we aim to set up a quantum kinetic theory in a rotational field. 

There are two editions for the quantum kinetic theory in the frame of Wigner function. One is the covariant version~\cite{degroot} for the Wigner function $W(x,p)$ defined in $8$-dimensional phase space, and the other is the equal-time version~\cite{BialynickiBirula:1991tx} for the Wigner function $W_0(x,{\bm p})$ in $7$-dimensional phase space. The advantage of the former is the explicit covariance under a Lorentz transformation, and the latter is directly related to the physical distributions defined in equal-time phase space. Of course, $W_0$ is not manifestly Lorentz covariant. In both the covariant and equal-time formalisms, an important aspect of the kinetic theory is that, the complex kinetic equation can be split up into a constraint and a transport equation, where the former is a quantum extension of the classical mass-shell condition, and the latter is a generalization of the Vlasov-Boltzmann equation. The complementarity of these two ingredients is essential for a physical understanding of quantum kinetic theory~\cite{zhuang,zhuang2,zhuang3}. In this paper, we focus on a complete description of the equal-time Wigner function in a rotational field, by considering the coupled constraint and transport equations. We will derive the classical and quantum transport equations for the two independent spin components, namely the number density and spin density, by using the semi-classical expansion of the kinetic equations.

The vortical field ${\bm \omega}$ of a system can be either generated self-consistently by the curl of the medium velocity ${\bm \omega}={\bm \nabla}\times{\bm v}$ or considered as an external field, depending on the particles we describe in kinetic equations. For light quarks which are constituents of the medium, the quark vorticity is just the rotation of the medium, but for heavy flavors which are considered as a probe of the medium, the vorticity in kinetic equations can be treated as an external field. We will consider in this paper the latter and neglect the collision terms among particles, in order to focus on the coupling between particles and the external rotational field. This version of the mean field approximation, which treats the particles quantum mechanically, but uses the classical approximation for the field, is widely used in electrodynamic kinetic theory~\cite{BialynickiBirula:1991tx,Hidaka:2016yjf,Huang:2018wdl,Gao:2018wmr,Weickgenannt:2019dks,Wang:2019moi,Wang:2020pej,kharzeev}. 

The paper is organized as follows. In section \ref{s2} we obtain in curved space the Dirac equation and its non-relativistic limit under a rotational field which is the basis to derive kinetic equations in the frame of Wigner function. We then calculate the kinetic equations for the covariant Wigner function $W(x,p)$ and its spin components in section \ref{s3}. By taking the energy integration of the covariant kinetic equations we obtain the constraint and transport equations for the spin components of the equal-time Wigner function $W_0(x,{\bm p})$ in section \ref{s4}. In section \ref{s5} we semi-classically solve the equal-time equations. We will focus on the coupling between the particle spin and the rotational field. We summarize the work in section \ref{s6}. 

\section{Dirac and Schr{\" o}dinger Equation in rotational field}
\label{s2}
The starting point to derive a relativistic or non-relativistic kinetic theory for quarks in Wigner function formalism is the Dirac equation or Schr\"odinger equation. The system under a rotational field can be equivalently regarded as a system at rest in a rotating frame, as has been discussed in Ref.\cite{Jiang:2016wvv} where the rotation of a quark system enhances the chiral symmetry restoration strongly and Ref.\cite{Liu:2018xip} where the covariant kinetic theory for chiral fermions in external electromagnetic field is extended to curved space systematically. To avoid confusion, we use in the following the indices $\{\mu,\nu,\lambda,\sigma\}$ and $\{\alpha,\beta,\gamma,\delta\}$ to separately describe Lorentz vectors and tensors in curved and flat space, known respectively as coordinate and non-coordinate basis~\cite{Nakahara:2003nw}.

The Lagrangian density for fermions under mean-field approximation in non-coordinate basis has the following form
\begin{equation}
\mathcal L=\sqrt{-g}\bar\psi\left(i\gamma^\alpha\partial_\alpha-m\right)\psi,
\end{equation}
where $\sqrt{-g}$ is related to the coordinate we choose. Considering that, in coordinate basis the tangent space $T_pM$ and cotangent space $T^*_pM$ are expanded in $\partial_\mu$ and $dx^\mu$, the coordinate transformation between the two spaces can be expressed as 
\begin{equation}
\hat e_\alpha=e_\alpha^{\ \mu}\partial_\mu,\ \ \ \ \ \ \ \ e_{\alpha\ }^{\ \mu} \in GL(m,\mathbb R), 
\end{equation}
where $\{\hat e_{\alpha}\}$ is required to be orthonormal with respect to $g\ (=g_{\mu\nu}dx^{\mu}\otimes dx^{\nu})$, which means the relation $g(\hat e_{\alpha},\hat e_{\beta})=e_{\alpha\ }^{\ \mu} e_{\beta\ }^{\ \nu}g_{\mu\nu}=\eta_{\alpha\beta}$ or inversely $g_{\mu\nu}=e^{\alpha\ }_{\ \mu} e^{\beta\ }_{\ \nu}\eta_{\alpha\beta}$. With the requirement of local Lorentz invariance, the Lagrangian density in coordinate basis becomes
\begin{equation}
\mathcal L=\sqrt{-g}\bar\psi\left[i\gamma^{\alpha}e_{\alpha\ }^{\ \mu}\left(\partial_{\mu}+\frac{i}{2}\Gamma^{\alpha\ \beta}_{\ \mu\ }\Sigma_{\alpha\beta}\right)-m\right]\psi
\end{equation}
with the affine connection $\Gamma^{\alpha\ \beta}_{\ \mu\ }=\eta^{\beta\gamma}e^{\alpha\ }_{\ \nu}(\partial_{\mu}e_{\ \gamma}^{\nu\ }+e^{\ \sigma}_{\gamma\ }\Gamma^{\nu}_{\ \mu\sigma})$.

We now consider a system under rotation with a constant vorticity denoted by ${\bm\omega}$. The local velocity of this rotating frame is given by ${\bf v}={\bm\omega}\times{\bm x}$, and the space-time metric is written as 
\begin{eqnarray}
\label{tetrad}
g_{\mu\nu} &=& \left(\begin{matrix}
1-{\bf v}^2 & -v_1   & -v_2 &  -v_3 \\
-v_1 & -1 & 0 & 0\\
-v_2  & 0 & -1 & 0\\
-v_3 & 0& 0 & -1
\end{matrix}\right),\nonumber\\
g^{\mu\nu} &=& \left(\begin{matrix}
1 & -v_1   & -v_2 &  -v_3 \\
-v_1 & -1+v_1^2 & v_1v_2 & v_1v_3\\
-v_2  & v_1v_2 & -1+v_2^2 & v_2v_3\\
-v_3 & v_1v_3& v_2v_3 & -1+v_3^2
\end{matrix}\right),
\end{eqnarray}
where we have introduced a specific tetrad~\cite{Jiang:2016wvv},  
\begin{equation}
e^{\alpha\ }_{\ \mu} = \delta^{\alpha\ }_{\ \mu}+\delta^{\alpha}_{i}\delta_{\mu}^0v_i,\ \ \ \ e_{\alpha\ }^{\ \mu} = \delta_{\alpha\ }^{\ \mu}-\delta_{\alpha}^0\delta_i^{\mu}v_i,
\end{equation}
and $\bm v=\bm\omega\times \bm r$ is the velocity of the coordinate transformation. It is worth noticing that, the choice of the tetrad is not unique, since the degrees of freedom of a $n$-dimensional metric is $(n+1)n/2$ and of the tetrad $n^2$. After plunging the chosen tetrad into the Lagrangian, we obtain
\begin{equation}
\label{lagrangian}
\mathcal L = \bar\psi\left[i\gamma^\mu\partial_\mu+\gamma_0{\bm \omega}\cdot\left({\bm x}\times(-i{\bm\nabla})+{\bm s}\right)-m\right]\psi
\end{equation}
with ${\bm s}=-\frac{1}{2}\gamma_0\gamma_5{\bm\gamma}=\frac{1}{2}\text{diag}({\bm\sigma},{\bm\sigma})$. Under the choice of the space-time metric (\ref{tetrad}), the higher orders of the rotational field, namely the terms $\sim {\bm \omega}^2$ and ${\bm \omega}^3$, vanish automatically, and only the linear term $\sim {\bm \omega}$ appears in the Lagrangian density. However, since the velocity of the coordinate transformation $\bm v=\bm\omega\times \bm r$ has been taken in the non-relativistic form, the Lagrangian (\ref{lagrangian}) is valid only for small ${\bm\omega}$, with $|\omega x| \ll 1$. From the structure of the Lagrangian, the rotational field ${\bm\omega}$ serves as a chemical potential coupled to the total angular momentum ${\bm J}={\bm x}\times \hat{\bm p}+{\bm s}$ which is conserved during the evolution of the system. 

With the known Lagrangian density it is easy to derive the Dirac equation for quarks in the rotational field, 
\begin{equation}
\label{dirac}
\left[i\gamma^\mu\partial_\mu+\gamma_0{\bm \omega}\cdot{\bm J}-m\right]\psi=0.
\end{equation}
The corresponding Schr\"odinger equation can be obtained by considering the non-relativistic limit of the Dirac equation in a standard way. Considering the stationary solution of the Dirac equation, $\psi(x)=\psi({\bm x})e^{-iEt}$, the stationary wave function $\psi({\bm x})$ satisfies the equation
\begin{equation}
\left[\left(\gamma_0{\bm \gamma}\cdot\hat{\bm p}-{\bm \omega}\cdot{\bm J}\right)+m\gamma_0\right]\psi=E\psi.
\end{equation}
To move to the familiar non-relativistic expression, we separate the quark energy into the mass and the kinetic energy, $E=m+\epsilon$, write the stationary wave function in a two-component form, $\psi({\bm x})=(\phi({\bm x}) ,\chi({\bm x}))^T$, and take the Pauli-Dirac representation for the $\gamma$-matrix, $\gamma_0=\left(\begin{matrix} I & 0 \\ 0 & -I \end{matrix}\right)$ and ${\bm\gamma}=\left(\begin{matrix} 0 & {\bm\sigma} \\ -{\bm\sigma} & 0 \end{matrix}\right)$, the two-component Dirac equation is then written as 
\begin{eqnarray}
&& \hat{\bm p}\cdot{\bm\sigma}\chi-{\bm\omega}\cdot{\bm J}\phi=\epsilon\phi,\nonumber\\
&& \hat{\bm p}\cdot{\bm\sigma}\phi-\left({\bm\omega}\cdot{\bm J}+2m\right)\chi=\epsilon\chi
\end{eqnarray}
with the total angular momentum ${\bm J}={\bm x}\times\hat{\bm p}+{\bm \sigma}/2$ in its two-dimensional form. From the second equation, the small component $\chi$ can be expressed as 
\begin{equation}
\chi = {\hat{\bm p}\cdot{\bm\sigma}\over 2m+\epsilon +{\bm\omega}\cdot{\bm J}}\phi \approx {\hat{\bm p}\cdot{\bm\sigma}\over 2m}\phi
\end{equation}
to the first order in $1/m$. Substituting it into the first equation leads to the Schr\"odinger equation for the large component $\phi$,
\begin{equation}
\label{schroedinger}
\left[{\hat{\bm p}^2\over 2m}-{\bm \omega}\cdot{\bm J}\right]\phi =\epsilon\phi
\end{equation}
which is the same as obtained by using non-relativistic Galilean transformation~\cite{anandan}.

To the second order in $1/m$, the small component $\chi$ becomes
\begin{equation}
\chi = {1\over 2m}\left(1-{\epsilon+{\bm\omega}\cdot{\bm J}\over 2m}\right)\hat{\bm p}\cdot{\bm \sigma}\phi,
\end{equation}
Taking the commutation relations between $x_i$ and $\hat p_j$ and between $\sigma_i$ and $\sigma_j$ and employing the above Schr\"odinger equation to the first order in $1/m$, we obtain the Scgr\"odinger equation to the second order,
\begin{equation}
\left[{\hat{\bm p}^2\over 2m}-{\hat{\bm p}^4\over 8m^3}-{\bm \omega}\cdot{\bm J}\right]\phi =\epsilon\phi,
\end{equation}
the only relativistic correction is to the kinetic energy. 

\section{Covariant kinetic equations}
\label{s3}
The core ingredient to describe the transport phenomena of a non-equilibrium system is the distribution function in phase space. Wigner function is the quantum analogue to the classical distribution function, and has been widely adopted in the investigation of quantum transport phenomena, such as spin polarization~\cite{Gao:2012ix,Wang:2020pej} in heavy ion collisions and pair creation in QED systems~\cite{BialynickiBirula:1991tx,zhuang,Sheng:2018jwf}. The covariant Wigner function $W(x,p)$ for fermions is defined as the ensemble average of the Wigner operator in vacuum state, and the Wigner operator is the four-dimensional Fourier transform of the covariant density matrix \cite{degroot}, 
\begin{equation}
\label{wigner}
W(x,p) = \int {d^4 y\over (2\pi)^4}\sqrt{-g(x)}e^{ip^\mu y_\mu}\langle\psi(x_+)U(x_+,x_-)\bar\psi(x_-)\rangle
\end{equation}
with $x_\pm =x\pm y/2$, where the gauge link is defined as
\begin{equation}
U(x_+,x_-)=e^{ig\int_{-1/2}^{1/2}ds y^\mu A_\mu(x+sy)}
\end{equation}
with the gauge field $A_\mu$. Since we focus in this paper on the rotational effect, we neglect the gauge field and in turn the gauge link. 

The covariant kinetic equation in the Wigner function formalism is derived by calculating the first-order derivatives of the density matrix and using the Dirac equations for the fields $\psi$ and $\bar\psi$. After a straightforward calculation, we obtain the equation of motion for the Wigner function in phase space which is equivalent to the equation of motion for the field in coordinate space, 
\begin{equation}
\label{kdirac}
\left[\gamma^\mu K_\mu + {\hbar\over 2} \gamma^5\gamma^\mu\omega_\mu-m\right]W(x,p)=0
\end{equation}
with the definitions of $K_\mu=\Pi_\mu + {i\hbar\over 2} D_\mu$ and $\omega_\mu=(0,{\bm \omega})$, where the extended momentum and derivative operators in phase space are defined as
\begin{eqnarray}
\Pi_\mu &=& (p_0+\pi_0, {\bm p}),\ \ \ \ \pi_0 = {\bm \omega}\cdot\left({\bm l}+{\hbar^2\over 4}{\bm \nabla}\times{\bm \nabla}_p\right)+\mu_B,\nonumber\\
D_\mu &=& (d_t, {\bm \nabla}),\ \ \ \ \qquad d_t = \partial_t-{\bm \omega}\cdot\left({\bm x}\times {\bm \nabla}+{\bm p}\times{\bm \nabla}_p\right)
\end{eqnarray}
with the orbital angular momentum ${\bm l}={\bm x}\times{\bm p}$. Note that, the rotational effect changes only the particle energy from $p_0$ to $p_0+\pi_0$ and time derivative from $\partial_t$ to $d_t$, and the vector momentum ${\bm p}$ and space derivative ${\bm \nabla}$ are not modified. In comparison with nuclear collisions at extremely high energy, the rotational effect will become more important in heavy ion collisions at intermediate energy where the baryon density becomes high. Aiming at a kinetic theory in such case, we have included here the baryon chemical potential $\mu_B$ which shifts the particle energy. In order to semi-classically solve the kinetic equations below, we have displayed the $\hbar-$dependence explicitly. It is clear that, the highest order quantum correction in the operators comes from the term $\sim \hbar^2$.     

Very different from the classical distribution which is a scalar function, the Wigner function in quantum case is a $4\times 4$ matrix in spin space, it includes in general case $16$ independent components. Because of their characteristic properties under Lorentz transformations, it is convenient to choose the $16$ matrices $1, i\gamma_5, \gamma_\mu, \gamma_\mu\gamma_5, \sigma_{\mu\nu}/2$ as the basis for an expansion of the Wigner function in spin space,  
\begin{equation}
\label{decomposition}
W={1\over 4}\left(F + i\gamma^5 P+\gamma^\mu V_\mu +\gamma^\mu\gamma^5 A_\mu+{1\over 2}\sigma^{\mu\nu} S_{\mu\nu}\right).
\end{equation}
All the components $\Gamma_\alpha=\{F, P, V_\mu, A_\mu, S_{\mu\nu}\}$ are real functions, since the basis elements transform under hermitian conjugation like the Wigner function itself, $W^+=\gamma_0 W \gamma_0$. They
can thus be interpreted as phase-space densities. Their physical meaning becomes clear in the equal-time formalism which will be discussed below.  

The expansion (\ref{decomposition}) decomposes the kinetic equation into $5$ coupled Lorentz covariant equations for the $5$ spinor components $\Gamma_\alpha$. Since these components are real and the operators $P_\mu$ and $D_\mu$ are self-adjoint, one can separate the real and imaginary parts of these $5$ complex equations,   
\begin{eqnarray}
\label{group1}
&& 2\Pi^\mu V_\mu + \hbar\omega^\mu A_\mu=2m F,\nonumber\\
&& \hbar D^\mu A_\mu=2m P,\nonumber \\
&& 4\Pi_\mu F -2\hbar D^\nu S_{\nu\mu}-\hbar\epsilon_{\mu\nu\alpha\beta}\omega^\nu S^{\alpha\beta}=4m V_\mu,\nonumber\\
&& -\hbar D_\mu P+\epsilon_{\mu\nu\alpha\beta}\Pi^\nu S^{\alpha\beta}-\hbar\omega_\mu F=2m A_\mu,\nonumber\\
&& \hbar(D_\mu V_\nu - D_\nu V_\mu)+2\epsilon_{\mu\nu\alpha\beta}\Pi^\alpha A^\beta+\hbar\epsilon_{\mu\nu\alpha\beta}\omega^\alpha V^\beta=2m S_{\mu\nu}
\end{eqnarray}
and 
\begin{eqnarray}
\label{group2}
&& \hbar D^\mu V_\mu  = 0,\nonumber\\
&& 2\Pi^\mu A_\mu+ \hbar\omega^\mu V_\mu=0,\nonumber \\
&& \hbar D_\mu F + 2\Pi^\nu S_{\nu\mu}-\hbar\omega_\mu P=0,\nonumber\\
&& 4\Pi_\mu P+\hbar\epsilon_{\mu\nu\alpha\beta}D^\nu S^{\alpha\beta}+ 2\hbar\omega^\nu S_{\mu\nu}=0,\nonumber\\
&& 2(\Pi_\mu V_\nu-\Pi_\nu V_\mu)-\hbar\epsilon_{\mu\nu\alpha\beta}D^\alpha A^\beta+ \hbar(\omega_\mu A_\nu-\omega_\nu A_\mu)= 0.
\end{eqnarray}
These equations are firstly derived by Vasak, Gyulassy and Eltz~\cite{vasak} for a QED system and are recently used to describe the chiral magnetic effect of a quark system in electromagnetic fields~\cite{Hidaka:2016yjf,Huang:2018wdl,Gao:2018wmr,Weickgenannt:2019dks,Wang:2019moi,Wang:2020pej}. These equations can further be divided into two groups. Those equations with terms multiplied by $p_0$ hidden in $P_0$ form the constraint group which links the Wigner function $W$ and its first order energy moment $p_0W$, and the others form the transport group which describes the evolution of $W$ in phase space. These will be discussed in more detail in the equal-time formalism.  

Similar to the Klein-Gordon equation for the wave function $\psi(x)$ which describes the plane-wave solution of the Dirac equation satisfying the on-shell condition $p^2-m^2=0$, we can obtain the phase-space version of the Klein-Gordon equation for the Wigner function $W(x,p)$ by acting the kinetic equation (\ref{kdirac}) with the operator $\gamma^\mu K_\mu+\hbar/2\gamma^5\gamma^\mu\omega_\mu+m$, which leads to
\begin{equation}
\label{kklein}
\left[K^\mu K_\mu-{i\over 2}\left[K_\mu, K_\nu\right]\sigma^{\mu\nu}-\hbar\gamma^5K^\mu\omega_\mu+{\hbar^2\over 4}\omega^\mu\omega_\mu-m^2\right]W(x,p)=0.
\end{equation}
We will see in the following that, this equation controls whether the particle is on the mass shell. 
 
\section{Equal-time kinetic equations}
\label{s4}
From the definition (\ref{wigner}), it is easy to see that the covariant Wigner function at a fixed time is related to the fields at all times. Therefore, the covariant kinetic equations in general case cannot be solved as an initial value problem, and we should go to the equal-time formalism of the kinetic theory, by doing energy integration of the covariant equations~\cite{zhuang}. The equal-time Wigner function is defined as 
\begin{equation}
\label{wigner0}
W_0(x,{\bm p}) = \int {d^3 {\bm y}\over (2\pi)^3}e^{-i{\bm p}\cdot{\bm y}}\langle\psi({\bm x}_+,t)U({\bm x}_+,{\bm x}_-,t)\psi^+({\bm x}_-,t)\rangle
\end{equation}
with ${\bm x}_\pm ={\bm x}\pm {\bm y}/2$. It is clear that, the equal-time Wigner function is not Lorentz covariant, and the two Wigner functions are related to each other through the energy integration, 
\begin{equation}
W_0(x,{\bm p}) =\int dp_0 W(x,p)\gamma_0.
\end{equation}
This indicates that, the equal-time Wigner function is the zeroth order energy moment of the covariant Wigner function. This is the reason why we label the equal-time Wigner function using the subscript $0$. In general case, particles moving in a medium are not on the mass shell, and the covariant Wigner function is equivalent to the collection of all the energy moments~\cite{zhuang3} 
\begin{equation}
W_n(x,{\bm p})=\int dp_0 p_0^n W(x,p)\gamma_0
\end{equation}
with $n=0,1,2,...$. Only in the quasi-particle approximation where particles are on the shell and the covariant Wigner function satisfies the on-shell condition $W(x,p)(p^2-m^2)=0$, the two Wigner functions are equivalent to each other. 

Similar to the covariant scenario, the equal-time Wigner function is decomposed into $8$ components in spin space, 
\begin{equation}
W_0 = {1\over 4}\left(f_0+\gamma_5 f_1-i\gamma_0\gamma_5 f_2+\gamma_0 f_3+\gamma_5\gamma_0{\bm \gamma}\cdot{\bf g}_0+\gamma_0{\bm \gamma}\cdot{\bf g}_1-i{\bm \gamma}\cdot {\bf g}_2-\gamma_5{\bf \gamma}\cdot {\bf g}_3\right),
\end{equation}
the equal-time components $f_i(x,{\bm p})$ and ${\bf g}_i(x,{\bm p})\ (i=0,1,2,3)$ are the zeroth order energy moments of the corresponding covariant components $\Gamma_\alpha(x,p)$. 

By taking $p_0-$integration of the covariant equations (\ref{group1}) and (\ref{group2}), one obtains two groups of equal-time kinetic equations, 
\begin{eqnarray}
\label{transport}
&& \hbar \left(d_t f_0 + {\bm \nabla}\cdot {\bf g}_1\right) = 0,\nonumber\\
&& \hbar \left(d_t f_1 + {\bm \nabla}\cdot {\bf g}_0\right) =-2mf_2,\nonumber \\
&& \hbar d_t f_2 +2{\bm p}\cdot{\bf g}_3 = 2mf_1,\nonumber\\
&& \hbar d_t f_3 - 2{\bm p}\cdot {\bf g}_2 =0,\nonumber\\
&& \hbar \left(d_t {\bf g}_0+{\bm \nabla}f_1\right)-2{\bm p}\times{\bf g}_1+\hbar{\bm\omega}\times{\bf g}_0=0,\nonumber\\
&& \hbar \left(d_t {\bf g}_1+{\bm \nabla}f_0\right)-2{\bm p}\times{\bf g}_0+\hbar{\bm\omega}\times{\bf g}_1=-2m{\bf g}_2,\nonumber\\
&& \hbar \left(d_t {\bf g}_2+{\bm \nabla}\times {\bf g}_3\right)+2{\bm p}f_3+\hbar{\bm\omega}\times{\bf g}_2=2m{\bf g}_1,\nonumber\\
&& \hbar \left(d_t {\bf g}_3-{\bm \nabla}\times {\bf g}_2\right)-2{\bm p}f_2+\hbar{\bm\omega}\times{\bf g}_3=0,
\end{eqnarray}
and 
\begin{eqnarray}
\label{constraint}
&& 2\int dp_0 p_0 F = \hbar{\bm \nabla}\cdot{\bf g}_2- 2\pi_0f_3 +2mf_0-\hbar{\bm\omega}\cdot{\bf g}_3,\nonumber\\
&& 2\int dp_0 p_0 P = -\hbar{\bm \nabla}\cdot{\bf g}_3-2\pi_0f_2 -\hbar{\bm\omega}\cdot{\bf g}_2,\nonumber\\
&& 2\int dp_0 p_0 V_0 = 2{\bm p}\cdot {\bf g}_1-2\pi_0f_0 +2mf_3-\hbar{\bm\omega}\cdot {\bf g}_0,\nonumber\\
&& 2\int dp_0 p_0 A_0 = -2{\bm p}\cdot{\bf g}_0+ 2\pi_0f_1+\hbar{\bm\omega}\cdot{\bf g}_1,\nonumber \\
&& 2\int dp_0 p_0 {\bf V} = \hbar{\bm \nabla}\times{\bf g}_0-2{\bm p} f_0+2\pi_0 {\bf g}_1-\hbar{\bm\omega}f_1,\nonumber\\
&& 2\int dp_0 p_0 {\bf A}=-\hbar{\bm \nabla}\times {\bf g}_1-2{\bm p}f_1+2\pi_0 {\bf g}_0+\hbar{\bm\omega}f_0-2m{\bf g}_3,\nonumber\\
&& 2\int dp_0 p_0 S^{0i}{\bf e}_i =\hbar{\bm \nabla}f_3-2{\bm p}\times{\bf g}_3 +2\pi_0 {\bf g}_2+\hbar{\bm\omega}f_2,\nonumber\\
&& \int dp_0 p_0 \epsilon^{ijk} S_{jk}{\bf e}_i ={\hbar}{\bm \nabla}f_2-2\pi_0 {\bf g}_3-2{\bm p}\times{\bf g}_2-{\hbar}{\bm\omega}f_3+2m{\bf g}_0.
\end{eqnarray}

The kinetic equations (\ref{transport}) and (\ref{constraint}) form, respectively, the transport and constraint groups. The former is an extension of the Boltzmann equation, it describes the phase-space evolution of the $8$ equal-time distributions in a rotational field. The latter is an extension of the on-shell condition $f(x,p)(p^2-m^2)=0$ associated to the Boltzmann equation. Since particles are generally not on the mass shell, the off-shell constraints cannot be neglected arbitrarily, and only the two groups together form a complete description of the quantum system. This is firstly pointed out by Zhuang and Heinz for a QED system~\cite{zhuang,zhuang2,zhuang3}. 

The constraints play a tremendous role in calculating some of the physical distributions. Let's consider the energy density as an example. From the energy-moment tensor,
\begin{equation}
T_{\mu\nu}(x) = i\left\langle\bar\psi(x)\gamma_\mu\partial_\nu\psi(x)-\bar\psi(x)\gamma^0\epsilon^{ijk}\delta_{i\mu}\omega_jx_k\partial_{\nu}\psi(x)\right\rangle,
\end{equation} 
the energy distribution in phase space is the first-order energy moment of the covariant component $V_0$,
\begin{equation}
\varepsilon(x,{\bm p})=T_{00}(x,{\bm p})=\int dp_0 p_0V_0(x,p).
\end{equation}
Without the constraints (\ref{constraint}) which links the zeroth- and first-order energy moments, there is no way to calculate the energy distribution in the frame of kinetic theory. With the help from the constraints (\ref{constraint}), $\epsilon$ is a combination of the equal-time spin components, 
\begin{eqnarray}
\label{energy}
\varepsilon &=&{\bm p}\cdot{\bf g}_1-\pi_0f_0+mf_3-{\hbar\over 2}{\bm \omega}\cdot{\bf g}_0,
\end{eqnarray} 
where the components $f_1, f_3, {\bf g}_0$ and ${\bf g}_1$ are controlled by the transport equations (\ref{transport}).
      
\section{Semi-classical expansion}
\label{s5}
The equal-time kinetic equations can directly be solved for some non-perturbative problems like pair production in electromagnetic fields~\cite{BialynickiBirula:1991tx,zhuang,Sheng:2018jwf}. As a systematical way the semi-classical expansion is widely used in covariant~\cite{Hidaka:2016yjf,Huang:2018wdl,Gao:2018wmr} and equal-time~\cite{BialynickiBirula:1991tx,zhuang,zhuang2,zhuang3,guo} kinetic theories for massive~\cite{Wang:2019moi} and massless~\cite{Hidaka:2016yjf,Huang:2018wdl,Gao:2018wmr} fermions. We discuss in this section the semi-classical expansion of the equal-time kinetic equations (\ref{transport}) and (\ref{constraint}). Considering the fact that, the rotational field appears only up to the second order in $\hbar$, the kinetic equations at zeroth, first and second order of $\hbar$ contain already all the quantum effects from the rotation on the phase-space distributions. 

We take the $\hbar$ expansion for the covariant and equal-time wigner functions $W(x,p)$ and $W_0(x,{\bm p})$ and the operator $\Pi_\mu$, 
\begin{eqnarray}
W &=& W^{(0)}+\hbar W^{(1)}+\hbar^2 W^{(2)}+\cdots,\nonumber\\
W_0 &=& W_0^{(0)}+\hbar W_0^{(1)}+\hbar^2 W_0^{(2)}+\cdots,\nonumber\\
\Pi_\mu &=& \Pi_\mu^{(0)}+\hbar^2\Pi_\mu^{(2)},\ \ \ \ \Pi_\mu^{(0)} = \left(p_0+{\bm \omega}\cdot{\bm l}+\mu_B,{\bf p}\right),\ \ \ \ \Pi_\mu^{(2)} = \left({\bm \omega}\cdot({\bm \nabla}\times{\bm \nabla}_p)/4,{\bf 0}\right).
\end{eqnarray}
Note that the other operator $D_\mu$ contains only the classical part. 

We first consider the Klein-Gordon equation (\ref{kklein}) at the zeroth order in $\hbar$,  
\begin{equation}
\left[\Pi^{(0)}_\mu\Pi^{(0)\mu}-m^2\right]W^{(0)}(x,p)=0.
\end{equation}
This is just the on-shell condition for classical particles with energy,
\begin{equation}
p_0=E_p^\pm=\pm\epsilon_p-({\bm \omega}\cdot{\bm l}+\mu_B)
\end{equation}
with $\epsilon_p=\sqrt{m^2+{\bm p}^2}$. Different from the kinetic theory for QED where the electromagnetic fields do not affect the free-particle shell~\cite{Hidaka:2016yjf,Huang:2018wdl,Gao:2018wmr,Weickgenannt:2019dks,Wang:2019moi,Wang:2020pej,zhuang}, the rotational field here changes the shell from $\epsilon_p$ to $E_p$ due to the interaction of the orbital angular momentum with the rotational field. The reason is clear: the electromagnetic fields ${\bf E}$ and ${\bf B}$ are derivatives of the gauge potential but ${\bm \omega}$ appears directly in the effective gauge potential ${\bm \omega}\times {\bm x}$~\cite{anandan}. The derivative leads to the appearance of ${\bf E}$ and ${\bf B}$ at least at the first order in $\hbar$, but ${\bm \omega}$ starts to contribute at the zeroth order.       

Considering the two elementary solutions of the classical Wigner function, corresponding to the positive and negative energies,
\begin{equation}
W^{(0)}(x,p) = W^{(0)+}(x,p)\delta(p_0-E_p^+)+ W^{(0)-}(x,p)\delta(p_0-E_p^-),
\end{equation}
the constraint equations (\ref{constraint}) reduce the number of independent spin components from $8$ to $2$. The independent components can be chosen to be $f_0$ and ${\bf g}_0$, and the others can be expressed in terms of them explicitly,
\begin{eqnarray}
\label{constraint0}
f_1^{(0)\pm}&=&\pm {1\over \epsilon_p}{\bm p}\cdot{\bf g}_0^{(0)\pm},\nonumber\\
f_2^{(0)\pm}&=&0,\nonumber\\
f_3^{(0)\pm}&=&\pm {m\over \epsilon_p}f_0^{(0)\pm},\nonumber\\
{\bf g}^{(0)\pm}_1&=&\pm {{\bm p}\over \epsilon_p}f_0^{(0)\pm},\nonumber\\
{\bf g}^{(0)\pm}_2&=&{1\over m}{\bm p}\times{\bf g}^{(0)\pm}_0,\nonumber\\
{\bf g}^{(0)\pm}_3&=&\pm{1\over m\epsilon_p}\left[\epsilon_p^2{\bf g}^{(0)\pm}_0-{\bm p}({\bm p}\cdot{\bf g}^{(0)\pm}_0)\right].
\end{eqnarray}

It is now the point to understand the physics of the spin components at quasi-particle level. Expressing the charge current and total angular momentum tensor in terms of the equal-time Wigner function, it is clear that the independent components $f_0$ and ${\bf g}_0$ are, respectively, the particle number density and spin density, and ${\bf g}_1$ is the number current density~\cite{BialynickiBirula:1991tx,zhuang}. Taking the classical relation $f_1={\bm p}/|{\bm p}|\cdot{\bf g}_0$ for massless fermions, $f_1$ can be interpreted as the helicity density. The components $f_3$ and $f_2$ describe the contribution from spontaneous chiral and $U_A(1)$ symmetry breaking of the system to the particle mass~\cite{guo}. From the non-relativistic limit ${\bf g}_3\to{\bf g}_0$ and the comparison of the term $-m/(2m){\bm\sigma}\cdot{\bm \omega}$ in the Schr\"odinger equation (\ref{schroedinger}) in rotational field for particles with effective charge $m$ with the term $-e/(2m){\bm \sigma}\cdot{\bf B}$ in the Schr\"odinger equation in QED for particles with charge $e$, ${\bf g}_3$ which is known as the magnetic moment density~\cite{BialynickiBirula:1991tx,Bargmann:1959gz} in electromagnetic fields can be understood as the rotational moment density. Considering the classical relation ${\bf g}_2={\bm p}\times{\bf g}_0/m$, ${\bf g}_2$ describes the spin property in the direction perpendicular to the particle momentum. Using the above classical relations, the energy density in quasi-particle approximation is simply expressed in terms of the number distributions with positive and negative energy,
\begin{equation}
\varepsilon(x,{\bm p}) = E_p^+f_0^{(0)+}(x,{\bm p})+ E_p^-f_0^{(0)-}(x,{\bm p}).
\end{equation}  

Since any derivative is multiplied by a factor of $\hbar$, the classical limit of the transport equations (\ref{transport}) cannot describe the phase-space evolution of the classical components but shows again some of the relations appeared already in the classical constraints (\ref{constraint0}). To describe the dynamical evolution of the equal-time Wigner function, we should go to the first order of the transport equations (\ref{transport}),    
\begin{eqnarray}
\label{transport1}
&& d_t f^{(0)}_0+ {\bm \nabla}\cdot {\bf g}^{(0)}_1 =0,\nonumber\\
&& d_t f^{(0)}_1+ {\bm \nabla}\cdot {\bf g}^{(0)}_0+2mf^{(1)}_2 =0,\nonumber \\
&& d_t f_2^{(0)}+ 2{\bm p}\cdot{\bf g}^{(1)}_3-2mf^{(1)}_1 =0,\nonumber\\
&& d_t f^{(0)}_3- 2{\bm p}\cdot{\bf g}^{(1)}_2 =0,\nonumber\\
&& d_t {\bf g}^{(0)}_0+{\bm \nabla} f^{(0)}_1-2{\bm p}\times{\bf g}^{(1)}_1+{\bm\omega}\times{\bf g}^{(1)}_0 =0,\nonumber\\
&& d_t {\bf g}^{(0)}_1+{\bm \nabla} f^{(0)}_0-2{\bm p}\times{\bf g}^{(1)}_0+{\bm\omega}\times{\bf g}^{(0)}_1+2m{\bf g}^{(1)}_2 =0,\nonumber\\
&& d_t {\bf g}^{(0)}_2+{\bm \nabla}\times {\bf g}^{(0)}_3+2{\bm p}f^{(1)}_3+{\bm\omega}\times{\bf g}^{(0)}_2 -2m{\bf g}^{(1)}_1 =0,\nonumber\\
&& d_t {\bf g}^{(0)}_3-{\bm \nabla}\times {\bf g}^{(0)}_2-2{\bm p}f^{(1)}_2+{\bm\omega}\times{\bf g}^{(0)}_3 =0.
\end{eqnarray}
Eliminating the first-order components $f_i^{(1)}$ and ${\bf g}_i^{(1)}$ by a simple algebra and taking into account the classical relations (\ref{constraint0}), we obtain the transport equations for the two independent components $f_0^{(0)}$ and ${\bf g}_0^{(0)}$, 
\begin{eqnarray}
\label{f0g00}
&& \left[\partial_t +\left(\pm{{\bm p}\over \epsilon_p}+{\bm x}\times {\bm \omega}\right)\cdot{\bm \nabla}-({\bm \omega}\times {\bm p})\cdot{\bm \nabla}_p\right] f_0^{(0)\pm}=0,\nonumber\\
&& \left[\partial_t +\left(\pm{{\bm p}\over \epsilon_p}+{\bm x}\times {\bm \omega}\right)\cdot{\bm \nabla}-({\bm \omega}\times {\bm p})\cdot{\bm \nabla}_p\right] {\bf g}_0^{(0)\pm}=-{\bm\omega}\times{\bf g}_{0}^{(0)\pm}.
\end{eqnarray}

The two equations are both in the Vlasov from. The particle velocity appeared in the free-streaming terms is modified by the rotation induced linear velocity ${\bm x}\times{\bm \omega}$, and the classical part of the rotational potential $-{\bm \omega}\cdot{\bm l}$ in the Dirac equation leads to a mean-field force (Coriolis force) $-{\bm \nabla}(-{\bm \omega}\cdot{\bm l})=-{\bm \omega}\times{\bm p}$. For the spin density ${\bf g}_0$, there is an extra term ${\bm \omega}\times{\bf g}_0$ indicating the spin-rotation interaction, similar to the term ${\bf B}\times{\bf g}_0$ in spinor QED. From the transport equations we obtain the equations of motion of the system,
\begin{eqnarray}
\dot{\bm x} &=& \pm \frac{\bm p}{\epsilon_p} + {\bm x}\times {\bf \omega},\nonumber\\
\dot{\bm p} &=& -{\bm \omega}\times {\bm p}.
\end{eqnarray}
Considering positive energy, the total force acting on the particles  
\begin{equation}
{\bf F} = \epsilon_p\ddot{\bm x}=-{\bm \omega}\times{\bm p}-\epsilon_p{\bm \omega}\times({\bm x}\times{\bm \omega})
\end{equation}
contains both the Coriolis force and centrifugal force. 

In order to investigate spin induced anomalous phenomena in a rotational field, one needs to go beyond the classical limit and derive quantum transport equations. To this end, we consider the Klein-Gordon equation (\ref{kklein}) again to see if quantum particles are still on a mass shell . At the first order in $\hbar$, the whole operator acting on the Wigner function becomes
\begin{equation}
iD^\mu\Pi_\mu^{(0)}-i/2({\bm \omega}\times{\bm p})\cdot[{\bm \gamma}, \gamma_0]+\gamma_5{\bm p}\cdot{\bm \omega},
\end{equation}
which is $\gamma$-matrix dependent. Therefore, there is no longer a common mass shell for all the spin components. To confirm this conclusion, we try the quasi-particle solution of the first-order constraint equations with an on-shell condition $W^{(1)}(x,p)(p_0^2-{\cal E}_p^2)=0$. Note that, if the quasi-particle ${\cal E}_p$ does exist, it should be different from the classical energy $E_p$ due to the modification by quantum fluctuations. However, the procedure fails. Massive particles cannot be on the shell when quantum effect is included. The case here is very different from the chiral limit where massless particles are always on a shell at any order of $\hbar$~\cite{Gao:2018wmr}. The second procedure we try is the spin-dependent on-shell condition $\Gamma_\alpha^{(1)}(x,p)(p_0^2-{\cal E}_{p\alpha}^2)=0$. This procedure fails again. Neither a common on-shell nor a component-dependent on-shell can be the solution of the constraint equations for massive fermions~\cite{Wang:2019moi}. The quantum effects in a general kinetic theory are essentially reflected in two aspects, one is the spin, and the other is the off-shell constraint.

Without the on-shell condition, the constraint equations (\ref{constraint}) at the first order in $\hbar$ becomes 
\begin{eqnarray}
\label{constraint1}
\pm 2\epsilon_p f^{(1)}_3+2\Delta E_{p3} &=& {\bm \nabla}\cdot{\bf g}^{(0)}_2+2mf^{(1)}_0-{\bm\omega}\cdot{\bf g}^{(0)}_3,\nonumber\\
\pm 2\epsilon_p f^{(1)}_2+2\Delta E_{p2} &=& -{\bm \nabla}\cdot{\bf g}^{(0)}_3-{\bm \omega}\cdot{\bf g}^{(0)}_2,\nonumber\\
\pm 2\epsilon_p f^{(1)}_0+2\Delta E_{p0} &=& 2{\bm p}\cdot{\bf g}^{(1)}_1+2mf^{(1)}_3- {\bm \omega}\cdot{\bf g}^{(0)}_0,\nonumber\\
\pm 2\epsilon_p f^{(1)}_1+2\Delta E_{p1} &=& 2{\bm p}\cdot{\bf g}^{(1)}_0-{\bm \omega}\cdot{\bf g}^{(0)}_1,\nonumber\\
\pm 2\epsilon_p {\bf g}^{(1)}_1 +2\Delta{\bf E}_{p1} &=& {\bm \nabla}\times{\bf g}^{(0)}_0+2{\bm p} f^{(1)}_0-{\bm\omega}f^{(0)}_1,\nonumber\\
\pm 2\epsilon_p {\bf g}^{(1)}_0 +2\Delta{\bf E}_{p0} &=& {\bm \nabla}\times{\bf g}^{(0)}_1+2{\bm p}f^{(1)}_1-{\bm\omega}f^{(0)}_0+2m{\bf g}^{(1)}_3,\nonumber\\
\pm 2\epsilon_p {\bf g}^{(1)}_2 +2\Delta{\bf E}_{p2} &=& -{\bm \nabla}f^{(0)}_3+2{\bm p}\times{\bf g}^{(1)}_3-{\bm\omega}f_2^{(0)},\nonumber\\
\pm 2\epsilon_p {\bf g}^{(1)}_3 +2\Delta{\bf E}_{p3} &=& {\bm \nabla}f^{(0)}_2-2{\bm p}\times{\bf g}^{(1)}_2-{\bm\omega}f_3^{(0)}+2m{\bf g}^{(1)}_0
\end{eqnarray}
with the energy shifts 
\begin{equation}
\label{shift}
\Delta E_{p\alpha} = 2\int dp_0 (p_0-E_p)\Gamma_\alpha^{(1)}.
\end{equation}

To close the equal-time constraint equations (\ref{constraint1}) which are related to the covariant components through the energy shifts (\ref{shift}), we need to consider the semi-classical expansion of the covariant kinetic equations (\ref{group1}) and (\ref{group2}). At classical level, the vector component is proportional to $\Pi_\mu^{(0)}$, and both the vector and axial-vector are on the mass shell,
\begin{eqnarray}
V_\mu^{(0)} &=& \Pi_\mu^{(0)}f^{(0)}\delta(\Pi_\rho^{(0)}\Pi^{(0)\rho}-m^2),\nonumber\\
A_\mu^{(0)} &=& g_\mu^{(0)}\delta(\Pi_\rho^{(0)}\Pi^{(0)\rho}-m^2),
\end{eqnarray} 
where $f(x,p)$ and $g_\mu(x,p)$ are arbitrary Lorentz scalar and vector distributions. After a straightforward but a little bit tedious algebra, the vector and axial-vector at first order in $\hbar$ can be decomposed into
\begin{eqnarray}
\label{va1}
V_\mu^{(1)} &=& \Pi_{\mu}^{(0)} f^{(1)}\delta(\Pi_\rho^{(0)}\Pi_{(0)\rho}-m^2)+\Pi_\mu^{(0)}\omega^\nu A_\nu^{(0)}\delta'(\Pi_\rho^{(0)}\Pi^{(0)\rho}-m^2),\nonumber\\
A_\mu^{(1)} &=& g_\mu^{(1)}\delta(\Pi_\rho^{(0)}\Pi^{(0)\rho}-m^2)+\Pi_\mu^{(0)}\omega^\nu V_\nu^{(0)}\delta'(\Pi_\rho^{(0)}\Pi^{(0)\rho}-m^2)-\omega_\mu\Pi^{(0)\nu} V_\nu^{(0)}\delta'(\Pi_\rho^{(0)}\Pi^{(0)\rho}-m^2),
\end{eqnarray}
where $\delta'$ means the derivative of the $\delta$-function. 

Taking together the first order transport and constraint equations (\ref{transport1}) and (\ref{constraint1}) for the equal-time components and (\ref{va1}) for the covariant components, we determine uniquely the energy shifts
\begin{eqnarray}
&&\Delta E^{\pm}_{p0}=-\frac{1}{2}{\bm\omega}\cdot {\bf g}_0^{(0)\pm},\nonumber\\
&&\Delta E^{\pm}_{p1}=\mp\frac{{\bm\omega}\cdot{\bm p}}{2\epsilon_p}f_0^{(0)\pm},\nonumber\\
&&\Delta E^{\pm}_{p2}=-{1\over 2m}{\bm\omega}\cdot({\bm p}\times{\bf g}_0^{(0)\pm}),\nonumber\\
&&\Delta E^{\pm}_{p3}=\mp{1\over 2m\epsilon_p}\left[\epsilon_p^2{\bm\omega}\cdot{\bf g}_0^{(0)\pm}-({\bm \omega}\cdot{\bm p}({\bm p}\cdot{\bf g}_0^{(0)\pm})\right],\nonumber\\
&&\Delta{\bf E}^{\pm}_{p0}=-{1\over 2\epsilon_p^2}\left[m^2{\bm\omega}+{\bm p}({\bm p}\cdot{\bf \omega})\right]f_0^{(0)\pm},\nonumber\\
&&\Delta{\bf E}^{\pm}_{p1}=\mp{{\bm \omega}\over 2\epsilon_p}{\bm p}\cdot{\bf g}_0^{(0)},\nonumber\\
&&\Delta{\bf E}^{\pm}_{p2}=\frac{{\bm p}^2({\bf p}\times{\bm\omega})}{2m\epsilon_p^2}f_0^{(0)\pm}\nonumber\\
&&\Delta{\bf E}^{\pm}_{p3}=\mp {1\over 2m\epsilon_p}\left[\epsilon_p^2{\bm \omega}-{\bm p}({\bm p}\cdot{\bm\omega})\right]f_0^{(0)\pm}.
\end{eqnarray}
All the energy shifts will disappear when the external field is turned off. The reason is clear that, without the coupling between the total angular momentum and the rotational field, particles will keep at the classical shell. Note that, there are two solutions for any energy shift, corresponding to the two classical shells $E_p^\pm$ or $\pm\epsilon_p$. 

The transport and constraint equations (\ref{transport1}) and (\ref{constraint1}) not only fix the quantum correction from the off-shell effect to the classical mass shell, but also reduce the number of independent spin components at quantum level. Again there are only two independent components. Similar to the classical limit, we can still choose the number density $f_0^{(1)}$ and spin density ${\bf g}_0^{(1)}$ as the independent components, and the others are determined by them self-consistently, 
\begin{eqnarray}
f_1^{(1)\pm}&=&\pm{1\over \epsilon_p}{\bm p}\cdot{\bf g}_0^{(1)\pm},\nonumber\\
f_2^{(1)\pm}&=&-{1\over 2m\epsilon_p^2}\left[\epsilon_p^2{\bm\nabla}\cdot{\bf g}_0^{(0)\pm}-({\bm p}\cdot{\bm\nabla})({\bm p}\cdot{\bf g}_0^{(0)\pm})\right],\nonumber\\
f_3^{(1)\pm}&=&\pm{m\over \epsilon_p}f_0^{(1)\pm}\mp{1\over 2m\epsilon_p}{\bm p}\cdot({\bm\nabla}\times{\bf g}_0^{(0)\pm}),\nonumber\\
{\bf g}^{(1)\pm}_1&=&\pm{{\bm p}\over \epsilon_p}f_0^{(1)\pm}\pm{1\over 2\epsilon_p}{\nabla}\times{\bf g}_0^{(0)\pm},\nonumber\\
{\bf g}_2^{(1)\pm}&=&{1\over m}{\bm p}\times{\bf g}_0^{(1)\pm}-{m\over 2\epsilon_p^2}{\bm\nabla}f_0^{(0)\pm}+{1\over 2m\epsilon_p^2}{\bm p}\times({\bm p}\times{\bm\nabla})f_0^{(0)\pm},\nonumber\\
{\bf g}^{(1)\pm}_3&=&\pm{1\over m\epsilon_p}\left[\epsilon^2_p{\bf g}_0^{(1)}-{\bm p}({\bm p}\cdot{\bf g}_0^{(1)})\right]\pm{{\bm p}\times{\bm\nabla}\over 2m\epsilon_p}f^{(0)\pm}_0+{1\over 2m\epsilon_p^2}\left[{\bm p}^2{\bm\omega}-{\bm p}({\bm p}\cdot{\bm \omega})\right]f_0^{(0)\pm}.
\end{eqnarray}
There are here three kinds of quantum corrections. The first one is a direct analogy to the classical relations shown in (\ref{constraint0}), by simply replacing the classical components $f_0^{(0)}$ and ${\bf g}_0^{(0)}$ by the first-order ones $f_0^{(1)}$ and ${\bf g}_0^{(1)}$. The second one comes from the derivative of the classical components, remembering that a derivative in kinetic equations is always accompanied by a factor of $\hbar$. The third correction is from the interaction with the external field which appears only in the rotational moment ${\bf g}_3$.      

The dynamical evolution of the equal-time Wigner function $W_0(x,{\bf p})$ at the first order in $\hbar$ is controlled by the transport equations (\ref{transport}) at the second order in $\hbar$,
\begin{eqnarray}
\label{transport2}
&& d_t f^{(1)}_0+ {\bm \nabla}\cdot {\bf g}^{(1)}_1 =0,\nonumber\\
&& d_t f^{(1)}_1+{\bm \nabla}\cdot{\bf g}^{(1)}_0+2mf^{(2)}_2=0,\nonumber \\
&& d_t f^{(1)}_2 +2{\bm p}\cdot{\bf g}^{(2)}_3-2mf^{(2)}_1 =0,\nonumber\\
&& d_t  f^{(1)}_3- 2{\bm p}\cdot {\bf g}^{(2)}_2=0,\nonumber\\
&& d_t {\bf g}^{(1)}_0+{\bm \nabla}f^{(1)}_1-2{\bm p}\times{\bf g}^{(2)}_1+{\bm\omega}\times{\bf g}^{(1)}_0 =0,\nonumber\\
&& d_t {\bf g}^{(1)}_1+{\bm \nabla}f^{(1)}_0-2{\bm p}\times{\bf g}^{(2)}_0+{\bm\omega}\times{\bf g}^{(1)}_1+2m{\bf g}^{(2)}_2 =0,\nonumber\\
&& d_t {\bf g}^{(1)}_2+{\bm \nabla}\times {\bf g}^{(1)}_3+2{\bm p}f^{(2)}_3+{\bm\omega}\times{\bf g}^{(1)}_2 -2m{\bf g}^{(2)}_1 =0,\nonumber\\
&& d_t {\bf g}^{(1)}_3-{\bm \nabla}\times {\bf g}^{(1)}_2-2{\bm p} f^{(2)}_2+{\bm\omega}\times{\bf g}^{(1)}_3=0.
\end{eqnarray}

By eliminating the second-order components and taking into account the classical and first-order kinetic equations (\ref{constraint0}), (\ref{transport1}) and (\ref{constraint1}), we obtain finally the transport equations for the two independent quantum distribution functions, namely the number density $f_0^{(1)}$ and spin density ${\bf g}_0^{(1)}$,
\begin{eqnarray}
\label{f0g01}
&&\left[\partial_t +\left(\pm {{\bm p}\over \epsilon_p} + {\bm x}\times {\bm \omega}\right)\cdot{\bm \nabla}-({\bm \omega}\times {\bm p})\cdot{\bm\nabla}_p\right] f_0^{(1)\pm}=0,\nonumber\\
&&\left[\partial_t +\left(\pm {{\bm p}\over \epsilon_p} + {\bm x}\times {\bm \omega}\right)\cdot{\bm \nabla}-({\bm \omega}\times {\bm p})\cdot{\bm\nabla}_p\right] {\bf g}_0^{(1)\pm}=-{\bm\omega}\times{\bf g}_{0}^{(1)}-{1\over 2\epsilon_p^4}{\bm p}\times({\bm p}\times{\bm\omega})({\bm p}\cdot{\bm \nabla})f_0^{(0)\pm}.
\end{eqnarray}
While the number density satisfies the same transport equation as the classical one, the coupling between the two independent components leads to a new term on the right-hand side of the quantum transport equation for the spin density.          

Following the way we used to derive transport equations in classical case and to the first order in $\hbar$, it is not a problem to obtain transport equations for the second-order components of the Wigner function. As has been mentioned above, the rotational field appears only up to the second order of $\hbar$ in the kinetic equations, there should be no more new information when going beyond the second order. 

\section{Summary and outlook}
\label{s6}
We investigated the quantum kinetic theory for a massive fermion system under a rotational field in Wigner function formalism. We derived the two groups of kinetic equations in covariant and equal-time versions, one is the constraint group which describes the off-shell effect in quantum case, and the other is the transport group which is the quantum analogy to the classical Boltzmann equation. For the structure of a quantum kinetic theory, the off-shell constraint is essentially important. It provides the physical interpretation for all the equal-time spin components, reduces the number of independent distribution functions, and closes the transport equations for the number density and spin density at classical level and quantum level. 

The interaction with the external rotational field through total angular momentum changes significantly the transport properties of the particles. The classical rotation-orbital coupling controls the dynamical evolution of the number distribution. It adds a linear velocity ${\bm x}\times{\bm \omega}$ to the particle velocity, and the induced Coriolis force ${\bm p}\times {\bm \omega}$ behaves as a mean field force acting on the particles. Apart from the classical coupling, the quantum rotation-spin coupling changes the spin distribution but does not affect the number distribution. While the two distributions are independent in classical limit, the number density influences the spin density at quantum level.        
   
There are still some questions we need to discuss in the future. One is the application of the obtained transport equations for the number and spin distributions. Since an equal-time transport equation can be solved as an initial value problem, the two transport equations can be used to describe the evolution of heavy quarks in high energy nuclear collisions to see the rotational effect on their propagation in hot medium. The collision terms should be included in a complete kinetic theory. The collisions among particles control the approaching of a system from non-equilibrium to equilibrium and will bring in the self-generated vorticity. The finite size effect is also important for a system under rotation. For a constant rotation, to guarantee the law of causality, the size of the system is under the constraint of $R_{max}\omega\le 1$. This means that, a rotational system should be finite, and therefore, one should consider the finite size effect on the Wigner function. 

{\bf Acknowledgement:} 
The work is supported by Guangdong Major Project of Basic and Applied Basic Research No. 2020B0301030008 and the NSFC under grant Nos. 11890712 and 12005112.

%--------------------------------------------------------------------------------------------------------------------------------

\end{document}